\documentclass[aps,prl,twocolumn,showpacs]{revtex4}
\usepackage{graphicx}

\begin{document}

\title{Doping Dependence of Superconductivity and Lattice Constants in Hole Doped $La_{1-x}Sr_xFeAsO$ }

\author{Gang Mu, Lei Fang, Huan Yang, Xiyu Zhu, Peng Cheng and Hai-Hu Wen}\email{hhwen@aphy.iphy.ac.cn }

\affiliation{National Laboratory for Superconductivity, Institute of
Physics and Beijing National Laboratory for Condensed Matter
Physics, Chinese Academy of Sciences, P. O. Box 603, Beijing 100190,
China}

\begin{abstract}
By using solid state reaction method we have fabricated the hole
doped $La_{1-x}Sr_xFeAsO$ superconductors with Sr content up to
0.13. It is found that the sharp  anomaly at about 150 K and the low
temperature upturn of resistivity are suppressed by doping holes
into the parent phase. Interestingly both the superconducting
transition temperature $T_c$ and the lattice constants (a-axis and
c-axis) increase monotonously with hole concentration, in sharp
contrast with the electron doped side where the $T_c$ increases with
a continuing shrinkage of the lattice constants either by dope more
fluorine or oxygen vacancies into the system. Our data clearly
illustrate that the superconductivity can be induced by doping holes
via substituting the trivalent La with divalent Sr in the LaFeAsO
system with single FeAs layer, and the $T_c$ in the present system
exhibits a symmetric behavior at the electron and hole doped sides,
as we reported previously.
\end{abstract}

 \maketitle

The discovery of superconductivity at 26$\;$K in iron-based layered
quaternary compound LaFeAsO$_{1-x}$F$_x$\cite{Kamihara2008} has
generated enormous efforts. This iron-based system attracts a lot of
attention since it may have an unconventional superconducting
mechanism as well as potential applications. Normally the Fe or Ni
elements form the materials with long range ferromagnetic order
which is detrimental to the spin singlet superconductivity, but it
seems to be an exception in this iron-based system. Many new high
temperature superconductors were then discovered and the
superconducting transition temperature has been quickly raised to
$T_\mathrm{c}=55\;$K in SmFeAsO$_{0.9}$F$_{0.1}$\cite{RenZA55K}. So
far most of the discovered superconductors are categorized into the
so-called electron doped side, while in La$_{1-x}$Sr$_{x}$FeAsO
superconductivity with $T_\mathrm{c}=25\;$K\cite{WenEPL} was
discovered by substituting La with Sr and assumed to be the first
hole doped superconductor in the iron-based system. Positive Hall
coefficient $R_H$ measured in these hole doped samples well
confirmed that the conduction is through hole-like charge carriers.
The discovery of the hole doped superconductors in the new system
puts a lot of constraints on the theoretical picture to understand
the superconducting mechanism. Later on superconductivity at about
38 K was found in Ba$_{1-x}$K$_x$Fe$_2$As$_2$\cite{Rotter}. This has
been repeated by many other groups by trying to replace the Ba site
with Sr and K site with Cs etc.\cite{Krellner,WangNL,ChuCW,ChenXH}
and the Hall effect data have also evidenced that they are hole
doped superconductors.

In this paper we present a systematic study on the evolution of the
superconductivity and the lattice constants with the content of Sr
in $La_{1-x}Sr_{x}FeAsO$. A close and intimate relationship between
the lattice constants and the superconducting transition temperature
has been found in the hole doped side: they both increase
monotonously with doped concentration of $Sr$. Our data and analysis
can also give an explicit explanation to the recent report by Wu et
al.\cite{ChenXH2} that they did not observe bulk superconductivity
in a nominally 15\% Sr doped as-grown sample
$La_{0.85}Sr_{0.15}FeAsO_{1-\delta}$. After having a detailed
analysis, we found that their so-called as-grown sample resided
actually at a Sr concentration of about $4\pm1$\%, far below 15\% in
the nominal composition as they believed. This prevents them to see
the superconductivity up to 26 K in a higher hole doping level. Our
results clearly illustrate that the bulk superconductivity in the
$LaFeAsO$ system can be induced by hole doping.

We employed a two-step solid state reaction method to synthesize the
$La_{1-x}Sr_xFeAsO$ samples. In the first step, $LaAs$ and $FeAs$
were prepared by reacting grains of La (purity 99.99\%), Fe powder
(purity 99.95\%) and As grains (purity 99.99\%) at 500 $^o$C for 8
hours and then 700 $^o$C for 10 hours. They were sealed in an
evacuated quartz tube when reacting. Then the resultant precursors
were thoroughly grounded together with Fe powder (purity 99.95\%),
$Fe_2O_3$ powder (purity 99.5\%) and $SrCO_3$ powder (purity 99.9\%)
in stoichiometry as given by the formula $La_{1-x}Sr_xFeAsO$.
Therefore our samples are not oxygen deficient from the starting
materials. The oxygen is even more than that in the stoichiometric
formula if counting the oxygen carried by $CO_2$. Sometime we also
use $SrO$ instead of $SrCO_3$ to make the superconductors
successfully. The obtained mixtures were pressed into pellets and
sealed in a quartz tube with high vacuum. Ar gas is not needed in
the quartz tube because $CO_2$ gas was produced during the reaction.
The materials were then heated up to 1150 $^o$C with a rate of 200
$^o$C/hr and maintained for 40 hours. Then a cooling procedure with
a rate of 100 - 200 $^o$C/hr was followed. We found that it is
necessary to bake the starting materials $Fe_2O_3$ and $SrCO_3$ at
350$^o$C for hours ahead of the synthesizing process in order to
remove the moisture absorbed by the powder. This baking seems very
essential to have samples with superconductivity when handling the
materials in a humid environment. We also found that using $Ta$ foil
to wrap the sample during the sintering should be strictly avoided
since it reacts with some components of the materials, probably with
$FeAs$ and $SrCO_3$. The $Ta$ foil will become dark and brittle if
it is used.

\begin{figure}
\includegraphics[width=8cm]{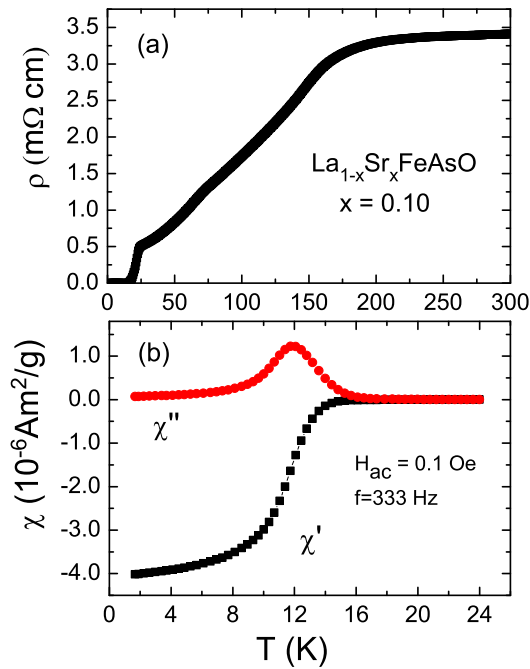}
\caption{(Color online) Temperature dependence of resistivity for
the $La_{0.9}Sr_{0.1}FeAsO$ sample. A flattening of resistivity in
high temperature part is obvious, which seems to be a common feature
of the hole doped iron-based superconductors. (b) The AC
susceptibility of the same sample measured with $H_{ac}$ = 0.1 Oe, f
= 333 Hz. } \label{fig1}
\end{figure}

\begin{figure}
\includegraphics[width=8cm]{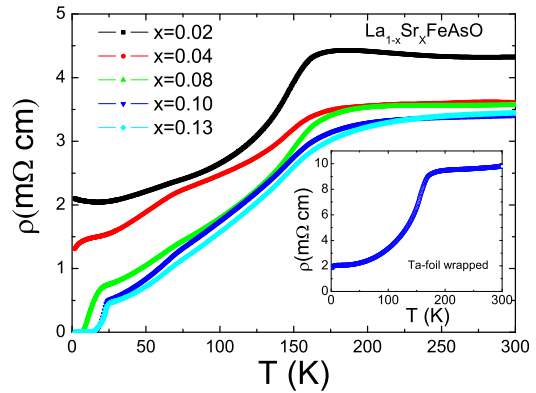}
\caption{Temperature dependence of resistivity of samples
$La_{1-x}Sr_{x}FeAsO$ with x = 0.02, 0.04, 0.08, 0.10, 0.13. One can
see that the resistivity anomaly and the low temperature upturn are
suppressed by doping more Sr into the system. The superconductivity
appears eventually. Inset: A typical resistivity curve measured on a
sample with nominal composition of 0.13, which was wrapped by a Ta
foil during the high temperature sintering (see text).} \label{fig2}
\end{figure}

In Fig.1(a) we present a typical set of resistive data for the
sample $La_{1-x}Sr_xFeAsO$ with x=0.1, it has an onset transition
temperature of about 25 K. Fig.1(b) shows the diamagnetic transition
measured on the same sample with AC susceptibility technique. An
estimation on the magnetic signal tells that the superconducting
shielding volume of the sample is beyond 60\%. Samples with this
kind of transitions can be easily repeated by carefully following
the synthesizing procedures we provided above. Our result is in
sharp contrast with that of Wu et al.\cite{ChenXH2}, they did not
see bulk superconductivity in $La_{1-x}Sr_xFeAsO$ with x=0.15, but
rather the resistivity exhibits a strange behavior in the normal
state with a tiny superconducting-like drop of resistivity at about
5 K. Our impression about their as-grown sample is that the true
doping content of Sr is far below the nominal composition x=0.15. In
addition, more complexity was involved with using the $Ta$ foil as
we addressed already. In the inset of Fig.2 we present a resistive
curve measured on the sample with the nominal composition of x=0.13
by using the $Ta$ foil. The curve looks very similar to the
so-called as-grown sample of Wu et al.\cite{ChenXH2}. Afterwards we
found out that the true Sr content in these samples are indeed very
low and they are still close to the undoped parent phase as
suggested by the large resistivity anomaly at about 150-160 K.

In the main frame of Fig.2 we show the resistivity data of our
samples made at various doping levels of Sr ranging from x=0.02 to
0.13. One can see that by using only 2\% doping of Sr, the upturn of
resistivity in the low temperature region has been clearly
suppressed, and the resistivity anomaly at about 150 K becomes
rounded compared to the parent phase.\cite{WangNLSDW} Up to 4\%
doping, both effects are suppressed strongly and a tiny resistivity
drop which may be due to superconductivity can be seen. At a doping
level of 8\%, a superconducting transition with onset point of 19.6
K occurs. At the doping level of 0.10-0.13, the superconductivity
becomes optimal with a onset transition temperature of 26 K. As
mentioned in our previous paper\cite{WenEPL}, one interesting
behavior for these hole doped samples is that the resistivity
anomaly at about 150 K is getting rounded, instead of being
suppressed completely as in electron doped samples. The similar
behavior appears in the recently discovered $(Ba,
Sr)_{1-x}K_x(FeAs)_2$ \cite{Rotter,WangNL,ChenXH,ChuCW} in which a
flattened behavior has been observed in the normal state resistivity
in high temperature region. We will show that this is an unique
feature for the hole-doped samples and associated very well with the
Hall coefficient data\cite{WenEPL}. Our present data obviously show
that by adding more Sr and thus holes into the sample, we can
actually tune the system away from the parent phase, and eventually
the superconductivity sets in. This is just like that in the double
layer system $Ba_{1-x}K_x(FeAs)_2$, introducing holes into the
sample by substituting Ba with K, the spin-density-wave like order
in the parent phase will be suppressed and finally superconductivity
wins out of the ground state.

In order to have a comprehensive understanding to the evolution
induced by the doping process, we have measured the X-ray
diffraction patterns for all samples. The lattice constants of
a-axis and c-axis are thus obtained. It is known that the lattice
constants (both a-axis or c-axis) will decrease with either fluorine
doping or more oxygen deficiency in
LaFeAsO\cite{RenZAOxygenVacancy}. This has been supposed to be an
effective way to increase $T_c$\cite{LaOFePHosono} in the iron-based
system. In Fig.3 we present the XRD patterns for five selected
samples. For clarity, the XRD data of samples with other doping
levels are not included here. It is clear that all main peaks of the
samples can be indexed to the tetragonal ZrCuSiAs-type structure,
leading to the determination of the lattice constants. An enlarged
view on the (102) peak for all samples clearly illustrate a
systematic evolution: the peak shifts monotonously to the left-hand
side direction in the order of 10 \% F-doped $LaFeAsO$, undoped
$LaFeAsO$ and then $La_{1-x}Sr_xFeAsO$. For the Sr doped
$La_{1-x}Sr_xFeAsO$ samples, the peaks also shift to more left-hand
side with more Sr concentration, indicating larger lattice
constants. This is reasonable since the radii of $Sr^{2+}$ is about
1.12 $\AA$, which is larger than that of $La^{3+}$ of about 1.06
$\AA$. The systematically expanded lattice constants with Sr doping
strongly suggest that Sr atoms go into the lattice structure of
LaFeAsO.

\begin{figure}
\includegraphics[width=8cm]{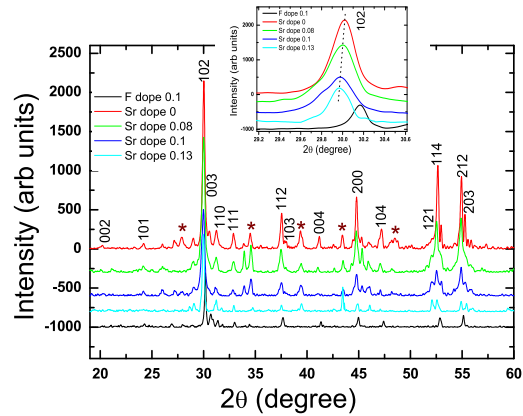}
\caption{(Color online) X-ray diffraction patterns for five selected
samples: 10 \% F-doped $LaFeAsO$, undoped $LaFeAsO$ and
$La_{1-x}Sr_xFeAsO$ with x = 0.08, 0.10 and 0.13. All main peaks can
be indexed to the tetragonal ZrCuSiAs-type structure. The asterisks
mark the impurity phase which may be the $FeAs$ and/or $Fe_2O_3$.
The inset shows an enlarged view of the (102) peak, it is clear that
the peak shifts monotonously to the left-hand side direction in the
order of 10 \% F-doped $LaFeAsO$, undoped $LaFeAsO$ and
$La_{1-x}Sr_xFeAsO$ with more doping. A dotted line through the peak
positions highlights this tendency. This indicates a gradual
expanding of the lattice constants. } \label{fig3}
\end{figure}

\begin{figure}
\includegraphics[width=8cm]{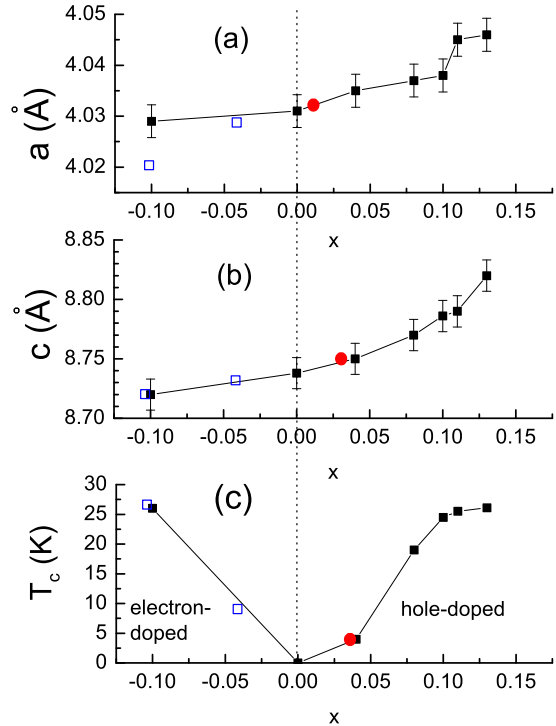}
\caption{(Color online) Doping dependence of (a) a-axis lattice
constant; (b) c-axis lattice constant; (c) the onset superconducting
transition temperature. Filled squares are data from our
measurements. For electron doped side, the variable "x" represents
the fluorine concentration or the similar electronic state by doping
oxygen vacancies. The red filled circles represent the lattice
constants and $T_c$ from the work of Wu et al.\cite{ChenXH2} if we
put their results on our general lines. The open squares roughly
tell the positions of the two samples of Wu et al.\cite{ChenXH2}
after 2 hour and 4 hour post-annealing in vacuum. One can see that
the lattice constant roughly show a monotonous variation versus
doping from electron doped region to hole doped one, but the
superconducting transition temperature $T_c$ shows a symmetric
behavior as we reported in our previous paper.\cite{WenEPL}. The
vertical dotted line indicates the undoped position.  } \label{fig4}
\end{figure}

Next we have a close look at the doping dependence of the lattice
constants and the superconducting transition temperature $T_c$, as
shown by Fig.4. It is clear that both the a-axis and c-axis lattice
constants expand monotonously starting from 10\% F-doped sample, the
undoped LaFeAsO sample, and the Sr doped samples. Interestingly the
onset superconducting transition temperature increases towards
either more hole doped or electron doped side. This strongly
suggests that the $T_c$ may have a symmetric behavior in both sides
as suggested in our earlier paper.\cite{WenEPL} If we put the
lattice constants from the samples of Wu et al.\cite{ChenXH2} onto
our plots, one can see that they reside quite close to the undoped
case. This is in consistent with the resistive transition curve
which shows a tiny drop of resistivity at about 5 K which may be
induced by the superconductivity in hole doped side. In addition,
this argument is further corroborated by the fact that their as
grown sample has a strong resistivity anomaly as the undoped sample,
together with the slight positive thermoelectric power, but still
negative Hall coefficient. If we further put their data points of
the vacuum annealed samples together, more interesting physics
emerges: the system was driven to the electron doped side again and
finally a superconductivity at about 26 K was observed with a sample
annealed for 4 hours. The lattice constants of these two annealed
samples follow also roughly well with the general tendency, i.e.,
they drop down when $T_c$ is increased towards more electron doping.
Above discussion indicate that the as-grown sample of Wu et
al.\cite{ChenXH2} is actually slightly hole doped although the
nominal Sr composition is 0.15. This prevents them to observe the
superconductivity up to 26 K in more hole doped samples as seen in
our experiments. However, an interesting point arises from their
annealed samples is that the system can be driven completely back to
the electron doped side by having the annealing in high vacuum and
at a high temperature.

To further support our argument, we have measured the Hall
resistance of the hole doped sample with x = 0.1 (the same one shown
in Fig.1). Fig.5 presents the data of Hall effect measurements. It
is clear that the Hall resistance is positive in wide temperature
region. An interesting discovery which was reported in our previous
paper\cite{WenEPL} as well as in the recent papers in $(Ba,
Sr)_{1-x}K_x(FeAs)_2$ is that the Hall coefficient $R_H$ has a huge
bump in the low temperature region, but it gradually becomes very
small and changes sign at a even high temperature. This seems to be
another common feature in hole doped iron-based superconductors.
When having a combined look at both Fig.2 and Fig.5, it is found
that the hump of Hall coefficient $R_H$ vanishes at about 200 K
where the normal state resistivity flattens out. This coincidence
may suggest that the two common features in the hole doped
iron-based superconductors have the same origin.  The positive Hall
coefficient $R_H$ as well as the close similarity with other
two-layer hole doped system $Ba_{1-x}K_x(FeAs)_2$ make our $Sr$
doped samples $La_{1-x}Sr_xFeAsO$ well footed on the category of
hole doped superconductors. Similar substitution with other rare
earth elements are worthwhile to try and may generate new
superconductors with higher $T_c$.

\begin{figure}
\includegraphics[width=8cm]{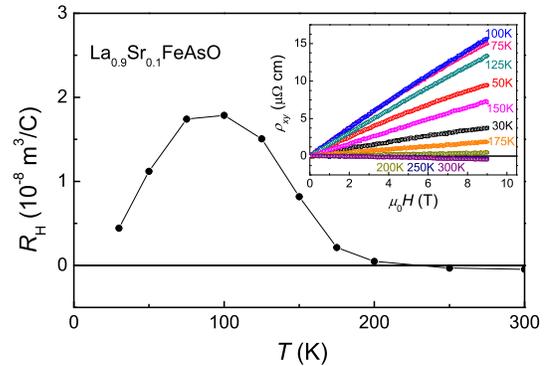}
\caption{(Color online) Temperature dependence of Hall coefficient
$R_\mathrm{H}$ determined on the sample $La_{0.9}Sr_{0.1}FeAsO$. A
huge bump appears in the low temperature region, which seems to be a
common feature for all hole doped superconductors. Inset: The raw
data of the Hall resistivity $\rho_{xy}$ at different temperatures.
} \label{fig5}
\end{figure}

In summary, a systematic evolution of superconductivity and the
lattice constants in hole doped $La_{1-x}Sr_xFeAsO$ has been
discovered. By doping more Sr into the parent phase $LaFeAsO$, the
low temperature upturn and the sharp anomaly at about 150 K of
resistivity are suppressed and the superconductivity eventually sets
in. The superconducting transition temperature as well as the
lattice constants increase monotonously with $Sr$ concentration up
to x = 0.13 where $T_c$ = 26 K. Together with the positive Hall
coefficient $R_H$ and the similar temperature dependence with other
hole doped iron-based superconductors, we conclude that the
superconductivity can be induced by either hole doping (through
substitution of La by Sr) or electron doping (through substitution
of oxygen by fluorine, or by inducing oxygen vacancies) in the
single layer system $LaFeAsO$.

We are grateful to Zidan Wang for fruitful discussions. This work is
supported by the Natural Science Foundation of China, the Ministry
of Science and Technology of China (973 project: 2006CB01000,
2006CB921802), the Knowledge Innovation Project of Chinese Academy
of Sciences (ITSNEM).

\end{document}